\documentclass{appolb}

\usepackage{graphicx}
\usepackage{amsmath}
\usepackage{citesort}

\begin{document}

\title{The 2D Mott-Hubbard transition \\ in presence of a parallel magnetic field}
\author{A. Avella and F. Mancini
\address{Dipartimento di Fisica ``E.R. Caianiello'' - Unit\`a INFM di
Salerno \\ Universit\`a degli Studi di Salerno, I-84081 Baronissi
(SA), Italy}}
\maketitle

\begin{abstract}
The half-filled two-dimensional Hubbard model in presence of a
uniform and static parallel magnetic field has been studied by
means of the Composite Operator Method. A fully self-consistent
solution, fulfilling all the constrains coming from the Pauli
principle, has been found. The relevant features of a
metal-insulator transition in presence of a magnetic field have
been analyzed. The results qualitatively agree with the ones
recently obtained by means of experimental investigations.
\end{abstract}


The response of a two-dimensional (2D) electronic system to a
parallel magnetic field is very intriguing and several anomalous
properties have been observed. There is a general agreement that
the observed behavior is related to the spin polarization, but
further studies, both theoretical and experimental, are needed. In
this paper we concentrate on the metal-insulator transition (MIT)
driven by a in-plane magnetic field. Recent experiments on
Si-MOSFET \cite{1} and GaAs \cite{2} have shown that by increasing
field the spin system polarizes and the system undergoes a MIT
before reaching the full polarization. Apparently, an important
role is played by the electron-electron interaction, being
$r_{s}=U/K$ (the ratio of Coulomb interaction energy to the mean
kinetic energy) very large.

In order to make a qualitative and preliminary study of this
phenomenon we consider the 2D Hubbard model in presence of a
parallel external magnetic field. Since a parallel field does not
couple to the orbital motion of electrons, the Hamiltonian is
given by
\begin{equation}
H=\sum_{\mathbf{ij}}\left( -4t\,\alpha _{\mathbf{ij}}-\mu \,\delta
_{\mathbf{ ij}}\right) c^{\dagger }\left( i\right) \,c\left(
j\right) +U\sum_{\mathbf{i} }n_{\uparrow }\left( i\right)
\,n_{\downarrow }\left( i\right) -\frac{1}{2}
h\sum_{\mathbf{i}}n_{3}\left( i\right)
\end{equation}
where $c\left(i\right)$ and $c^{\dagger}\left(i\right)$ are the
annihilation and creation operators of electrons in spinorial
notation; $i=\left( \mathbf{i},t\right)$ where $\mathbf{i}$ are
vectors of a 2D Bravais lattice; $\mu$ is the chemical potential;
$\alpha_{\mathbf{ij}}$ denotes the projector on first-neighbor
sites; $U$ is the local Coulomb interaction, $n_{\sigma }\left(
i\right) =c_{\sigma }^{\dagger }\left( i\right) \,c_{\sigma
}\left( i\right) $ is the charge density of the electrons with
spin $\sigma $; $n_{3}\left( i\right) $ is the third component of
the spin density operator; $h$ is proportional to the intensity of
the external magnetic field. In the framework of the Composite
Operator Method (COM) \cite{3}, we introduce the basis $\psi
^{\dagger }(i)=\left( \xi ^{\dagger }(i),\eta ^{\dagger
}(i)\right) $ where $\xi (i)=\left( 1-n(i)\right) c(i)$ and $\eta
(i)=n(i)\,c(i)$ are the Hubbard operators responsible for the
transitions $\left| 0\right\rangle \leftrightarrow \left| \sigma
\right\rangle $ and $\left| \sigma \right\rangle \leftrightarrow
\left| \uparrow \downarrow \right\rangle $, respectively. The
composite operator $\psi (i)$ satisfies the equation of motion
\begin{equation}
\mathrm{i}\frac{\partial }{\partial t}\psi \left( i\right)= \left(
\frac{1}{2}h\,\sigma _{3}-\mu \right) \psi \left( i\right)
-2t\left( 1+\tau _{3}\right) c^{\alpha }\left( i\right)
+\frac{1}{2}U\left( 1-\tau _{3}\right) \eta \left( i\right)
-4t\,\tau _{3}\,\pi \left( i\right)
\end{equation}
where $\vec{\sigma}$ acts on the spin degree of freedom $\sigma
=\uparrow $, $\downarrow $ and $\vec{\tau}$ on the internal degree
of freedom $\psi =\xi$ , $\eta $. $\vec{\sigma}$ and $\vec{\tau}$
are Pauli matrices. We also use the notation $\phi ^{\alpha
}\left( \mathbf{i},t\right) =\sum_{\mathbf{j} }\alpha
_{\mathbf{ij}}\phi \left( \mathbf{j},t\right) $. Moreover, we have
$\pi (i)=\frac{1}{2}\sigma ^{\mu }\,n_{\mu }(i)\,c^{\alpha
}(i)+\xi (i)\left[ c^{\dagger \alpha }(i)\,\eta (i)\right] $ where
$\sigma _{\mu }=(1,\vec{\sigma})$, $\sigma ^{\mu
}=(-1,\vec{\sigma})$ and $n_{\mu }(i)=c^{\dagger }(i)\,\sigma
_{\mu }\,c(i)$ describe the total charge- $\left( \mu =0\right) $
and spin- $\left( \mu =1\text{, }2\text{, } 3\right) $ density
operators.


In the polar approximation \cite{3} we linearize the equation of
motion by projecting the source on the basis $\psi (i)$. Then, the
retarded Green's function $S(\mathbf{k},\omega
)=\mathcal{F}\left\langle \mathcal{R}\left[ \psi (i)\,\psi
^{\dagger }(j)\right] \right\rangle $, where $\mathcal{F}$ and
$\mathcal{R}$ are the Fourier transform and the usual retarded
operators, respectively, has the following expression
\begin{equation}
S(\mathbf{k},\omega )=\sum_{l=1}^{4}\frac{\sigma ^{\left( l\right)
}\left( \mathbf{k}\right) }{\omega -E^{\left( l\right) }\left(
\mathbf{k}\right) + \mathrm{i}\,\delta }
\end{equation}
where the energy spectra $E^{\left( l\right) }\left(
\mathbf{k}\right) $ are the eigenvalues of the energy matrix
$\varepsilon \left( \mathbf{k}\right) =\mathcal{F}\left\langle
\left\{ J\left( \mathbf{i},t\right) ,\psi ^{\dagger }\left(
\mathbf{j},t\right) \right\} \right\rangle I^{-1}(\mathbf{k})$ and
the spectral density matrices $\sigma ^{\left( l\right) }\left(
\mathbf{k }\right) $ are calculated by means of the formula
$\sigma _{\alpha \beta }^{(l)}(\mathbf{k})=\Omega _{\alpha
l}(\mathbf{k} )\sum_{\gamma }\Omega _{l\gamma
}^{-1}(\mathbf{k})\,I_{\gamma \beta }( \mathbf{k})$ where $\Omega
(\mathbf{k})$ is the matrix whose columns are the eigenvectors of
the energy matrix $\varepsilon \left( \mathbf{k}\right) $ and
$I(\mathbf{k })=\mathcal{F}\left\langle \left\{ \psi \left(
\mathbf{i},t\right) ,\psi ^{\dagger }\left( \mathbf{j},t\right)
\right\} \right\rangle $ is the normalization matrix. The explicit
expressions of $E^{\left( l\right) }\left( \mathbf{k}\right) $ and
$\sigma ^{\left( l\right) }\left( \mathbf{k} \right) $ will be
given elsewhere. Calculations show that the Green's function
depends on the following set of parameters: $\mu $, $m$, $\Delta
_{\sigma }$, $p_{\sigma }$. $m=\frac{1}{2}\left\langle n_{3}\left(
i\right) \right\rangle$ is the magnetization per site. The
parameters $\Delta _{\sigma }$ and $p_{\sigma }$ describe a
constant shift of the bands and a band width renormalization,
respectively, and are defined as
\begin{align}
\Delta _{\sigma }& =\left\langle \xi _{\sigma }^{\alpha }\left( i\right)
\,\xi _{\sigma }^{\dagger }\left( i\right) \right\rangle -\left\langle \eta
_{\sigma }^{\alpha }\left( i\right) \,\eta _{\sigma }^{\dagger }\left(
i\right) \right\rangle \\
p_{\sigma }& =\frac{1}{4}\left[ \left\langle n_{\mu }^{\alpha }\left(
i\right) \,n_{\mu }\left( i\right) \right\rangle +2\left( -\right) ^{\sigma
}\left\langle n^{\alpha }\left( i\right) \,n_{3}\left( i\right)
\right\rangle \right] -\left\langle \left[ \xi _{\uparrow }\left( i\right)
\,\eta _{\downarrow }\left( i\right) \right] ^{\alpha }\eta _{\downarrow
}^{\dagger }\left( i\right) \,\xi _{\uparrow }^{\dagger }\left( i\right)
\right\rangle
\end{align}

The determination of these parameters is very crucial and wrong
results are easily obtained as shown in Ref.~\cite{4}. The
parameters $m$ and $\Delta _{\sigma }$ are expressed in terms of
the Green's function as $m = \frac{1}{2}\left(
C_{44}-C_{22}\right)$, $\Delta _{\uparrow } = C_{11}^{\alpha
}-C_{22}^{\alpha }$ and $\Delta _{\downarrow } = C_{33}^{\alpha
}-C_{44}^{\alpha }$. We have defined the correlation matrices
$C=\left\langle \psi \left( i\right) \,\psi ^{\dagger }\left(
i\right) \right\rangle $ and $C^{\alpha }=\left\langle \psi
^{\alpha }\left( i\right) \,\psi ^{\dagger }\left( i\right)
\right\rangle $. The other parameters $\mu $ and $p_{\sigma }$ are
not determined by the equation of motion and are fixed by choice
of the representation where the Green's functions are realized
\cite{5}. In the COM we choose the representation by requiring
that all the relations among the operators dictated by the algebra
(Pauli principle) are conserved also at the level of expectation
values. In the present study, this requirement leads to $C_{11} =
C_{33}$ and $C_{12} = C_{34} =0$. Because we are interested in the
study of the MIT, we consider the special case of half filling
($n=\left\langle n\left( i\right) \right\rangle=1$) where: $\mu
=\frac{U}{2}$, $\Delta _{\uparrow } = -\Delta _{\downarrow }$,
$p_{\uparrow } = p_{\downarrow }-2m$ and $C_{12} \equiv C_{34}
\equiv 0$. It is worth to note that these latter relations are a
manifestation of the particle-hole symmetry which is conserved
owing to the choice of the representation. Any other choice of the
representation will lead, in the context of the
pole-approximation, to a violation of the symmetry \cite{4}.
Finally, we have a set of three coupled self-consistent equations
which determine the three parameters which are left: $m$, $\Delta
=\Delta _{\uparrow }$, $ p=p_{\uparrow }$.

\begin{figure}[t]
\begin{center}
\includegraphics[width=0.49\textwidth]{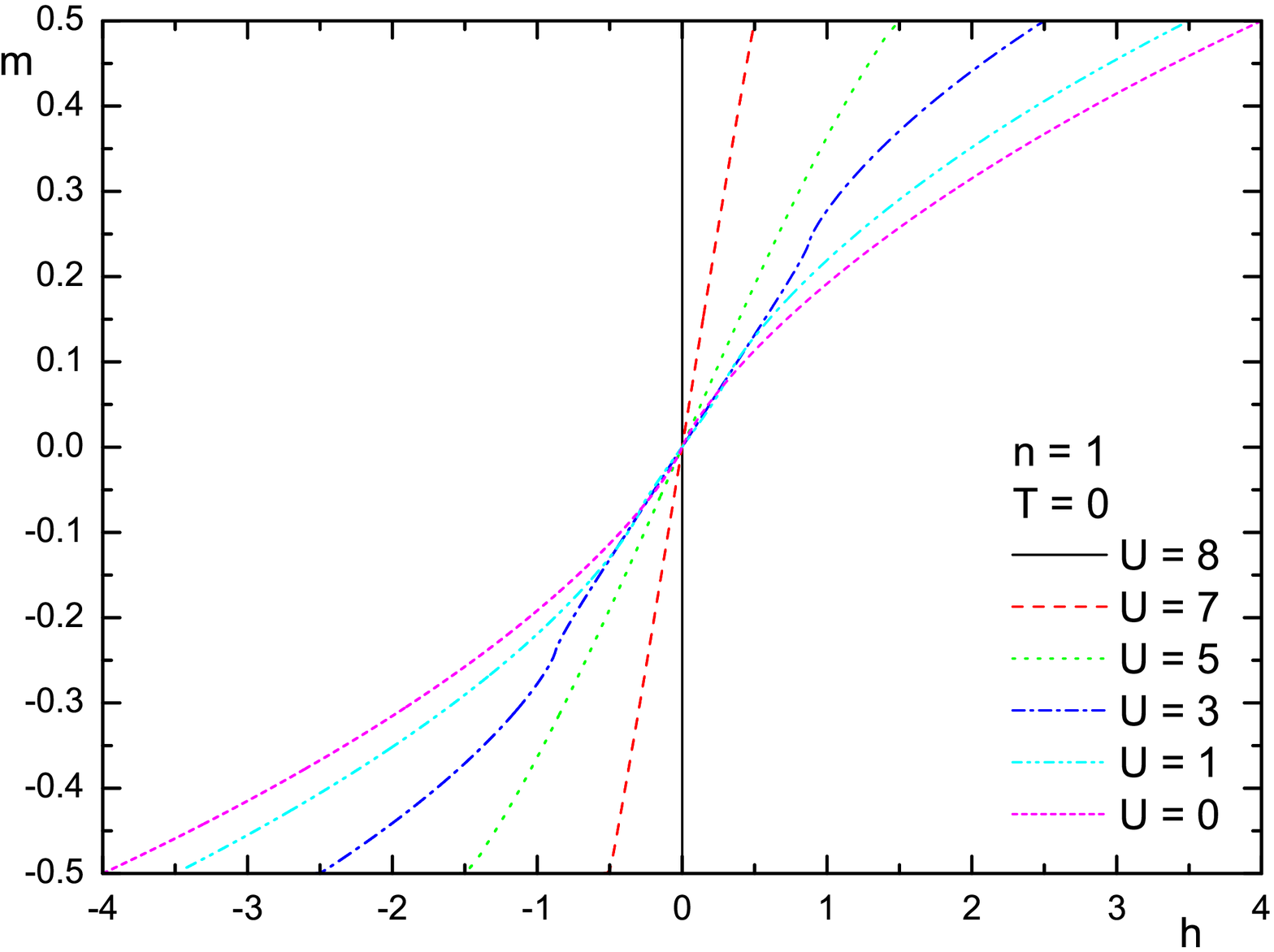}
\hfill
\includegraphics[width=0.49\textwidth]{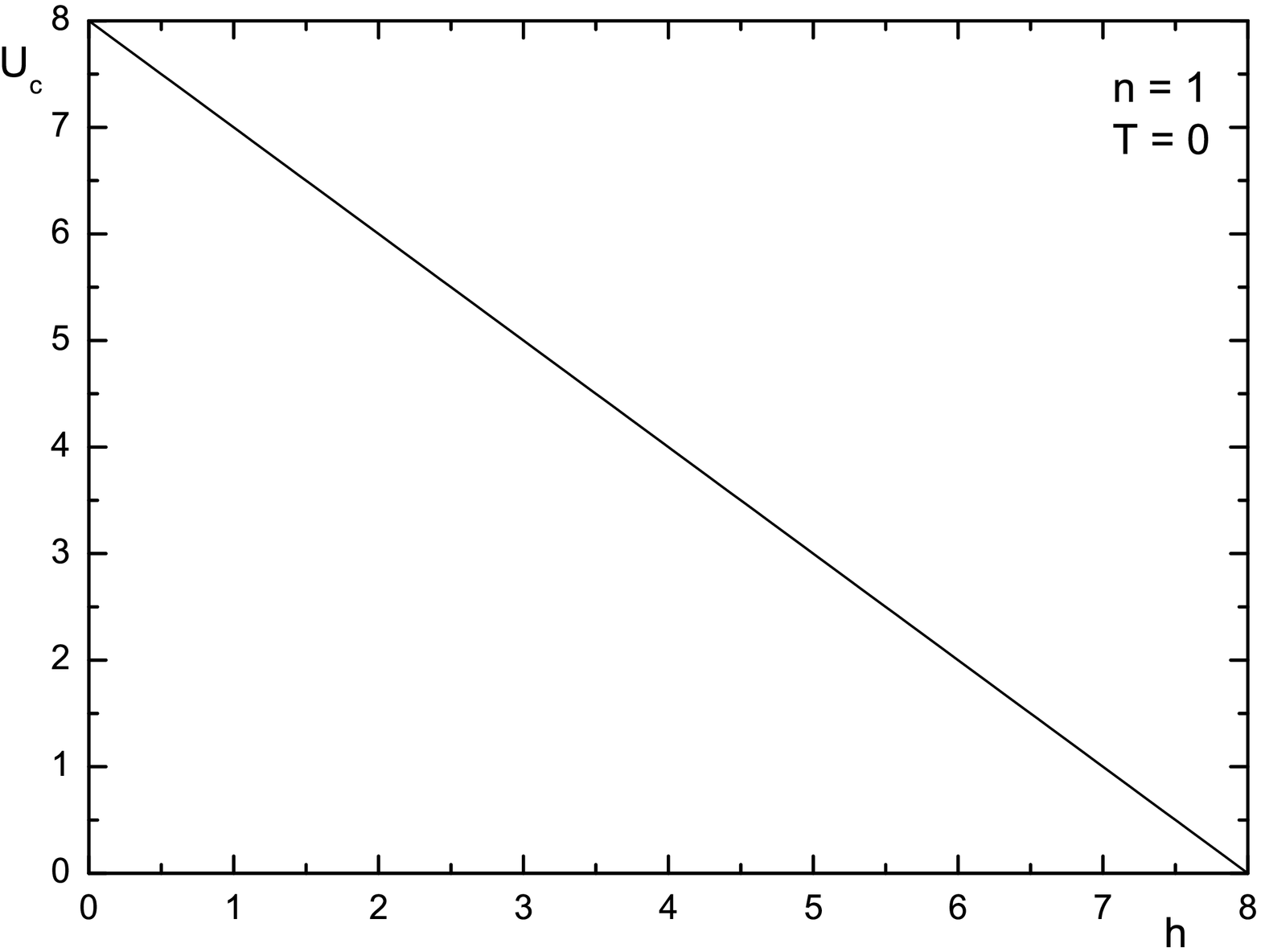}
\end{center}
\caption{(left) The magnetization $m$ as a function of the
external magnetic field $h$ for $T=0$, $n=1$ and various values of
the Coulomb repulsion $U$; (right) The critical value $U_c$ of the
Coulomb repulsion $U$ for the MIT as a function of the external
magnetic field $h$ for $T=0$ and $n=1$. $t$ is taken as unity.}
\label{Fig}
\end{figure}

In Fig.~\ref{Fig} (left panel) we plot the magnetization $m$
versus the magnetic field $h$. The magnetization is an increasing
function of both the applied magnetic field and the Coulomb
interaction $U$. It reaches the saturation value (i.e., $1/2$) at
a critical value of the magnetic field, which depends on the
intensity of the Coulomb interaction. At zero temperature $T=0$,
when $U$ approaches the bandwidth $W_{\textrm{2D}}=8t$, the
magnetization experiments a discontinuous jump from zero up to the
saturation value. We have also inspected the analytical behavior
of the static susceptibility by analyzing the self-consistent
equations in the limit of very low magnetic fields. Results show a
divergence when the Hubbard repulsion approaches the bandwidth at
zero temperature. The double occupancy decreases when increasing
both the interaction and the magnetic field. The latter provides
the spins of the electrons with an orientation and, due to the
Pauli principle, reduces the double occupancy. There is a quite
good agreement between COM results and Gutzwiller ones \cite{6}.

The MIT can be studied by looking at the density of states (DOS):
the opening of a gap in the DOS is a signal of the transition from
metallic to insulating phase. In Ref.~\cite{8} we have studied the
MIT exhibited by the Hubbard model in absence of magnetic field
for the 2D and 3D cases. It was found that the transition is
driven by the Coulomb interaction: there is a critical value $U_c$
where the MIT occurs. In particular, the value $U_c=1.68W$
($W_{\textrm{2D}}=8t$ and $W_{\textrm{3D}}=12t$ for the 2D and 3D
system, respectively) was reported. In presence of a magnetic
field the value of $U_c$ is drastically influenced. In
Fig.~\ref{Fig} (right panel) we plot the critical value versus the
magnetic field at zero temperature. As we turn on a rather small
magnetic field, the critical value $U_c$ suddenly jumps from
$U_c=1.68W$ to $U_c=W$. This discontinuity at zero field is
related to the discontinuity of the magnetization, as shown in
Fig.~\ref{Fig} (left panel). By increasing $h$, $U_c$ decreases
and vanishes when the field equates the bandwidth at zero
temperature (i.e., $U_c(h,T=0)=W-h$), in qualitative agreement
with the experimental findings.

In conclusion, our study shows that the 2D Hubbard model in
presence of a parallel magnetic field can describe the
experimental evidence of a field-driven MIT. The transition is
controlled by the field and disappears for some critical value of
it. A more detailed discussion of the MIT and of the order
parameter controlling the transition will be reported elsewhere.

\end{document}